\documentclass[amsmath,amssymb,floatfix,prb,epsf,showpacs,twocolumn,graphicx]{revtex4}
\usepackage{amsmath,amssymb,natbib,bm,graphicx,url,epsfig}
\usepackage[ansinew]{inputenc}
\usepackage{bm}

\newcommand{\be}{\begin{equation}}
\newcommand{\ee}{\end{equation}}
\newcommand{\bea}{\begin{eqnarray}}
\newcommand{\eea}{\end{eqnarray}}
\newcommand{\bes}{\begin{equation}\begin{split}}
\newcommand{\ees}{\end{split}\end{equation}}

\newcommand{\tw}{\bar{\omega}}

\newcommand{\bb}{\beta}

\begin{document}

\title{Pseudogap in underdoped cuprates and spin-density-wave fluctuations}
\author{Tigran A.~Sedrakyan$^{1,2}$ and Andrey  V.~Chubukov$^1$}
 \affiliation{$^1$Department of Physics, University of Wisconsin-Madison, Madison,
WI 53706, USA\\
$^2$Department of Physics, University of Maryland, College Park, MD  20742, USA}

\date{\text{January 31, 2010}}
\begin{abstract}
We analyze fermionic spectral function in the spin-density-wave (SDW) phase of quasi-2D cuprates at small but finite $T$. We use a non-perturbative approach and sum up infinite series of thermal self-energy terms, keeping at each order
nearly-divergent $(T/J) |\log \epsilon|$ terms, where $\epsilon$ is a deviation from a pure 2D, and neglecting regular $T/J$ corrections. We show that, as SDW order decreases,  the spectral function in the antinodal region acquires peak/hump structure: the coherent peak  position scales with SDW order parameter, while the incoherent hump remains roughly at the same scale as at $T=0$ when SDW order is the strongest. We identify the hump with the pseudogap observed in ARPES and argue that the presence of coherent  excitations at low energies gives rise to magneto-oscillations in an applied field. We show that the same peak/hump structure appears in the density of states and in optical conductivity.
\end{abstract}
\pacs{71.10.Hf, 75.10.Jm, 74.25.Dw}

\maketitle

\section{Introduction}

Understanding of the phase diagram of cuprate superconductors continue to
 be one of central topics in theoretical condensed matter physics.~\cite{mike}
Parent compounds of cuprates are quasi-2D antiferromagnetic
insulators, heavily overdoped cuprates are Fermi liquids.
 In between,  systems are $d-$wave superconductors at low $T <T_c$
 and display the pseudogap  behavior at larger $T_c < T< T^*$.
How an insulator
 transforms into a Fermi liquid and what is the origin of the pseudogap
are still the subjects of intensive debates among researchers.

The pseudogap region exists both in underdoped and overdoped cuprates, but
 the physics evolves substantially between these two limits. For
 overdoped cuprates, there is rather strong evidence~\cite{ali_over} that the pseudogap region
 is best described as a disordered superconductor, when the gap is already developed but the phase coherence is not yet set.~\cite{kiv,levin,millis_franz,randeria}
 In this doping range, fermions are reasonably well described as strongly interacting quasiparticles with a large, Luttinger-type underlying
 Fermi surface (FS).~\cite{acs}
  The $d-$wave
pairing in this doping range most naturally originates from the exchange of overdamped collective bosonic excitations of which spin-fluctuation mediated pairing is the key candidate.~\cite{acs,scal,pines,manske,
Kyung:2009}
In underdoped cuprates, situation is more complex. On one hand,
  ARPES data taken at low energies (below $50 meV$) and low $T$
were interpreted as the indication
 that the underlying FS still has Luttinger form, and the gap extracted
 from the position of the still visible narrow peak in the spectral function
has a simple $d-$wave, $\cos 2 \phi$ form over the whole FS, including antinodal region around $(0,\pi)$ and symmetry-related points.~\cite{camp_latest,arpes_pgap}
 On the other hand,  ARPES data taken in the antinodal region show that
the spectral function in the pseudogap regime develops a broad maximum
 at around $100-200meV$.~\cite{arpes_pgap,shen,he_2}
  The jury is still out~\cite{discussion} whether the observed high-energy hump and low-energy peak are separate features,
 or the peak and the hump describe the same gap, $\Delta (k)$, which strongly deviates from $\cos 2\phi$ form with underdoping.
The experimental results in Refs. \onlinecite{arpes_pgap,else,arcs,shen,two_gaps,he,he_2} were interpreted both ways.
 We side with the idea that the pairing gap remains $\cos 2\phi$ even in underdoped materials, and the hump is a separate feature, associated with
 Mott physics. We further take the point of view that the origin of the hump is
 the development of precursors to
 a Heisenberg-type antiferromagnetically ordered state at
half-filling.~\cite{rep,JS,vilk,tremblay,bansil,sad_2,
Vilk:1995,borejsza:2004}
  These precursors
 are generally termed as SDW precursors though one should keep in mind that
the half-filled state is the strong coupling version of SDW and is best described by the Heisenberg model with short-range exchange interaction.
The SDW precursor scenario has been wildly discussed
 in mid-90th,~\cite{schraiman,rep,vilk,
tremblay,
LTP:2006} and is near-universally accepted scenario for electron-doped cuprates~
\cite{tremblay,bansil,Motoyama:2007,millis_opt}. For hole-doped cuprates, it was, however,  put aside for a number of years
 in favor of
 non-Fermi liquid type scenarios.~\cite{nfl_pg}
The SDW scenario, however, re-gained support in the last few years,
 after magneto-oscillation experiments in a  field of $30-60T$
 detected long-lived Fermi liquid quasiparticles near small electron and hole FSs.~\cite{qo}  Such FS geometry is expected for an SDW ordered state,~\cite{rep}
 and early theory prediction was that a field drives the system
towards a SDW instability.\cite{zhang} Long-range  antiferromagnetic order
 in applied field has been explicitly detected
 in recent neutron-scattering experiments on underdoped $YBCO$ (Ref.~\onlinecite{hinkov}). [Another widely discussed scenario of quantum oscillations, which we will not consider here, is a $d-$wave density-wave order.~\cite{sudip}]

\begin{figure}[t]
\centerline{\includegraphics[width=75mm,angle=0,clip]{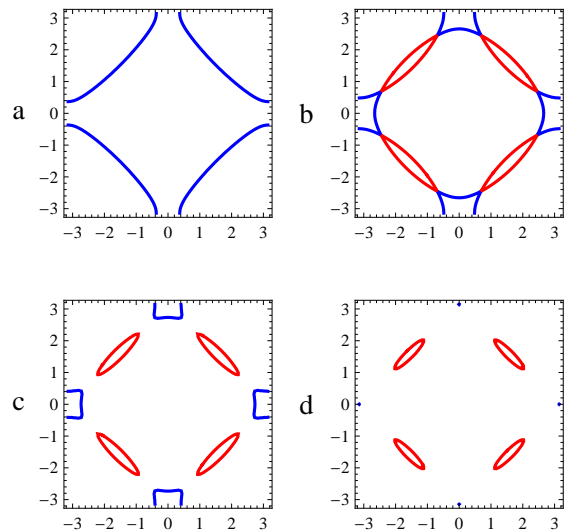}}
\caption{(Color online) Evolution of the FS with increasing SDW
order. (a) -- paramagnetic phase;  (b) -- SDW order is about to develop
 ($\langle S_z\rangle=0^{+}$). The shadow FS emerges, but the residue of fermionic excitations at the shadow FS is $0^{+}$; (c) -- small $\langle S_z\rangle$, both hole and electron pockets are present; (d) -- a larger  $\langle S_z\rangle$, only hole pockets  around $(\pi/2,\pi/2)$ remain.
} \label{FSS}
\end{figure}

In this communication, we analyze the consistency between the description of quantum oscillations {\it and} the pseudogap in underdoped cuprates within SDW scenario. The problem is the following: to explain quantum oscillations
 one has to assume the existence of small electron pockets.~\cite{norm_mill,granath}  Such pockets do exist in the SDW scenario near $(0,\pi)$ and symmetrey-related points, but they are present only if SDW order
$\langle S_z (Q)\rangle = \langle S_z\rangle$ is smaller than a threshold ($Q = (\pi,\pi)$).  For larger $\langle S_z\rangle$,
  only hole pockets around $(\pi/2,\pi/2)$ are present, while
  excitations near $(0,\pi)$  have a gap of order $4t^{\prime} \sim 0.2 eV$ (see Fig.~\ref{FSS}).  Antinodal pseudogap detected in ARPES experiments  in zero field
 is of the same magnitude.~\cite{arpes_pgap}
 A field of $40-60T$ is too small to affect energies of  $0.2 eV$,
 hence the same $200 meV$
pseudogap should be present in the ordered SDW state~\cite{mike_private}.
If this pseudogap is viewed as a precursor to SDW, one could expect that
 it simply sharpens up in the ordered SDW phase and transforms into the true
antinodal gap. But then there will be no electron pockets
 in the SDW phase, in disagreement with  magneto-oscillation experiments.
  The pseudogap and quantum oscillations can be reconciled within SDW scenario
 only if the evolution of the spectral function between paramagnetic and SDW states is more complex than just the sharpening of the pseudogap, and antinodal spectral function in the SDW phase contains both, high-energy pseudogap and electron pockets.  This co-existence also explains ARPES data in the superconducting state~\cite{camp_latest,arpes_pgap} because pairing of coherent fermions near electron pockets gives rise to a sharp peak in the spectral function at the gap energy.

To address this issue, we consider how
 fermionic Green's function $G(k, \omega)$  evolves within SDW phase,
as SDW order gets smaller. We depart from Heisenberg antiferromagnet with exchange interaction $J = 4t^2/U$ and
consider analytically how $G(k, \omega)$ is affected by
 thermal fluctuations which in quasi-2D systems destroy long-range order already at $T \ll J$. We neglect regular $T/J$ corrections but sum up infinite series of self-energy terms which contain  powers of $\beta = (T/\pi J) |\log \epsilon|$, where $\epsilon$
is a parameter which
measures deviations from pure two-dimensionality and which we will just  use as a lower cutoff of logarithmically divergent 2D integrals. In practice, $\epsilon$ is e.g., the ratio of the hoppings
 along $z$-axis and in $xy$ plane~\cite{Dare:1996}.
 SDW order disappears when $\beta = \beta_{cr} = O(1)$. This yields a set of integral equations for the Green's function and $\langle S_z\rangle$ which we obtain and solve.

In the terminology of Ref.~\onlinecite{sub-ch}, our computations are valid in the  renormalized-classical regime of quasi-2D systems.
The idea that, in this regime, at low enough $T$, one cannot restrict with Eliashberg or FLEX approximations and has to include self-energy and vertex corrections on equal footings has been put forward in Refs. \cite{rep,vilk}.
The computational procedure that we are using is
 similar to eikonal approximation in the scattering theory. Such
 procedure has been used in the study 1D CDW systems
 by Sadovskii~\cite{sad} and others~\cite{sad_1}, and has been applied to cuprates in Ref.~\onlinecite{JS}  to analyze SDW precursors in the paramagnetic
phase (for latest developments, see Ref. \onlinecite{sad_2}).
  Our computation has one advantage over earlier works:  in the SDW-ordered state
we don't need to assume that $T$ is larger than some threshold $T_0$ to restrict
 with only thermal fluctuations (i.e., with the contributions from zero Matsubara frequency). All we need is a small $\epsilon$ such that $T/J |\log \epsilon| = O(1)$ even when $T/J$ is small. In a paramagnetic phase, eikonal approximation is only valid when $T >T_0$, and $T_0$ increases as one moves away from the SDW phase. We assume that near SDW boundary
$T_0$ is small and apply our theory also to a paramagnetic phase.
  Our results for a paramagnet are in full agreement with Ref.~\onlinecite{JS}.

Note that in our theory (and in Ref. \onlinecite{JS}), the paramagnetic state is a Fermi liquid at  the lowest energies at $T=0$. We don't discuss here a possibility that a new, non-Fermi liquid state emerges near the region where SDW order is lost~\cite{sachdev3}.  We also don't discuss possibilities of more complex spin order and of open electron Fermi surfaces~\cite{norm_mill,granath}.
Our results are expected to survive if SDW order is incommensurate (with
${\bf Q}$ still near $(\pi,\pi)$, but  whether our results survive if the system develops  a stripe order  remains to be seen.

\begin{figure}[t]
\centerline{\includegraphics[width=85mm,angle=0,clip]{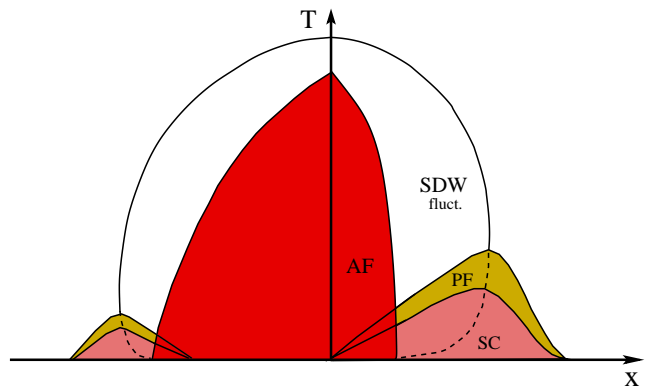}}
\caption{(Color online) Schematic phase diagram of hole and electron-doped cuprates.
The regions of antiferromagnetism (AF),
superconductivity (SC) and pairing fluctuations (PF) are shaded.
In the PF region, there are vortex excitations and
 large Nernst signal.~\protect\cite{nernst}
 Precursors to SDW appear at a non-zero T due to strong thermal fluctuations.
 In the region where SDW precursors are already developed,
 the onset temperature for the pairing increases with increasing $x$ (Ref.~\protect\onlinecite{Sachdev1}),
 in the region with no SDW precursors, it decreases with increasing $x$ (Ref.~\protect\onlinecite{acs}), the maximum $T_c$ is in the area between the two regimes. A similar phase diagram has been proposed in Ref.~\protect\onlinecite{Sachdev2}.}
\label{phase-space}
\end{figure}

We found that the spectral function $A(k, \omega) = (1/\pi) |Im G(k, \omega)|$
 near $(0,\pi)$ in the SDW state has a peak and a hump.
Both originate from a single peak  at the value of the $T=0$
SDW gap at $(0,\pi)$.
The hump moves little as SDW order decreases and just gets broader, while
 the peak follows $\langle S_z\rangle$, shifts to lower energies as SDW order decreases, and vanishes  $\beta = \beta_{cr}$,
when the system enter the paramagnetic phase.
  At $\beta \geq \beta_{cr}$, only the hump (the pseudogap) remains, and the spectral function at antinodal $k=k_F$ has camel-like structure, with a minimum at $\omega =0$. As $\beta$ increases further,
 $A(k_F, \omega =0)$ increases and eventually the spectral function at $k =k_F$
 develops a  single-peak at $\omega=0$, as it should be for a system with a large, Luttinger FS.

Rewinding this backwards, from a paramagnet to an SDW state,
  we see that the system first develops a pseudogap as a precursor to
SDW. When SDW order sets in,
 the pseudogap  sharpens up, but, in addition,
there also appears a true quasiparticle peak at low-energies. The residue of the peak increases as $\langle S_z\rangle$ increases.
 When  $\langle S_z\rangle$  is below the threshold,  electron pockets are present,
 and the spectral function near $(0,\pi)$ has a low-energy coherent peak
 and a  hump at about the same energy as the pseudogap in a paramagnetic phase.
 When SDW order gets larger, electron pockets eventually disappear, peak and hump come closer to each other and merge when $\langle S_z\rangle$ reaches its maximum.

This peak/hump structure also shows up in the density of states and in
the optical conductivity $\sigma (\omega)$. In the Mott-Heisenberg limit
($2U \langle S_z\rangle$ is larger than free-fermion bandwidth)
the conductivity at $T=0$ is zero up to
 a charge-transfer gap $U \sim 1.7 eV$. Once SDW order gets smaller, the peak at $U$ splits into a hump which slowly shifts to a higher frequency, and a
 peak whose energy  scales as $2U \langle S_z\rangle$.  In addition, there appears a metallic Drude component at the smallest frequencies.  This behavior is quite consistent with the measured $\sigma (\omega)$ in electron-doped cuprates,
 where SDW phase extends over a substantial doping range.~\cite{Onose04,Motoyama:2007,millis_opt,Armitage09}

We also found that, at a finite $T$, the system  in the pseudogap phase
 retains the memory about pockets. There are no real pockets in the sense that
 there is no two-peak structure of the spectral function at zero frequency  along zone diagonal, but we found that, when $\beta \geq \beta_{cr}$, the spectral weight at $\omega =0$ extends almost all the way between the original FS at $k_F$ and the ``shadow'' FS  at $k = Q - k_F$ (see Fig.~\ref{spectral-nodal}).
 As $\beta$ becomes larger, the $k-$range where the spectral weight is finite shrinks, and at large $\beta$ the spectral function recovers Drude-like
 structure typical for a metal with a large, Luttinger FS.

This analysis can be extended into a superconducting state. A system does
 not need to possess coherent quasiparticles to develop a pairing instability,~\cite{acf,mike_artem} but fermionic coherence emerges below the actual
$T_c$ much in the same way as it emerges in the SDW ordered state. The spectral function in the antinodal region then displays a coherent superconducting peak
 and a hump centered at, roughly, the energy of the antinodal
SDW gap at $T=0$.  This picture is consistent with the
data from Refs.~\onlinecite{arpes_pgap,else}.

The overall conclusion of our analysis is the phase diagram of the cuprates presented in Fig.~\ref{phase-space}. A similar phase diagram has been proposed in Ref.~\protect\onlinecite{Sachdev2}.
 At $T\neq 0$, there is a region
where the system displays SDW precursors. In this region, magnetic excitations are moderately damped, propagating magnons, and  magnetically-mediated $d-$wave
pairing interaction decreases as doping decreases
due to a reduction of the electron-magnon vertex.~\cite{Sachdev1,oleg,schr}
In this region,
the antinodal  pseudogap, caused by SDW precursors, and a $d-$wave
 pairing gap co-exist.  Outside this region, SDW precursors do not emerge, and arcs and other pseudogap
features are caused by thermal fluctuations of a pairing gap.~\cite{millis_franz,arcs,ch_norm}

We discuss the computational procedure in the next section,
 present the results in Sec.~3 and conclusions in Sec.~4.

\section{Computational procedure}

Our point of departure is the mean-field, SDW
 theory~\cite{SWZ,Chubukov92,zlatko_singh,rep}  of the
antiferromagnetically ordered state in  the large $U$ quasi-2D
Hubbard model at $T=0$. We assume $t-t^{\prime}$ dispersion in the XY
plane [$\epsilon_k = -2t (\cos k_x + \cos k_y) - 4 t^{\prime} \cos k_x \cos k_y$]
and weak dispersion along the $Z$ axis, which we will not keep explicitly in the formulas. Mean-field description neglects quantum fluctuations and is rigorously justified when the model is extended to $2S \gg 1$ fermionic flavors,~\cite{mus}
 but  qualitatively it remains
valid even for $S=1/2$ ($\langle S_z\rangle$  becomes $0.32$ instead of $0.5$).

  Long-range antiferromagnetic order splits the fermionic dispersion into valence and conduction bands, separated by $U$, and gives rise to a  two-pole
 structure of the bare fermionic Green's function:
\begin{eqnarray}
\label{twopoles}
G_0\left(\omega,{\bf k}\right)=u_k^2 G^c_0 + v^2_k G^v_0,
\end{eqnarray}
where  $u_k, v_k =\sqrt{(1 \mp 4 t \gamma_k/E_k)/2}$,
 $\gamma_k=(\cos k_x+\cos k_y)/2$, and
\begin{equation}
G^c_0 = \frac{1}{\omega-{E}^c_k},~~G^v_0 = \frac{1}{\omega-{E}^v_k}.
\end{equation}
The dispersions of conduction and valence
electrons are given by
${E}_k^{c,v}=\pm E_k-4t^{\prime}\cos k_x \cos k_y -\mu$, where
$E_k=\left[\Delta^2_0 +16 t^2 \gamma^2_k\right]^{1/2}$,
 $\Delta_0 = U\langle S_z\rangle$, and $\mu \approx -\Delta_0$ is the chemical potential.

The shape of the FS depends on the value of $\Delta_0$. We show the evolution of the FS with increasing $\Delta_0$ in Fig.~\ref{FSS}.
For small $\Delta_0$ both hole and electron pockets are present, for larger $\Delta_0$ only hole pockets remain.
At large $U/t$, which we assume to hold,  valence and conduction bands are well separated near half-filling at $T=0$, $u_k^2\approx v_k^2\approx 1/2$, and the FS only contains hole pockets.

The value of $\langle S_z\rangle$ is determined by the self-consistency condition
\be
\langle S_z\rangle = \int \frac{d^2 k}{(2\pi)^2} u_k v_k \int \frac{d \omega}{\pi} n_F (\omega) Im \left[G^c_0 - G^v_0\right],
\label{ch_1}
\ee
where both $G^{c}_0$ and $G^{v}_0$ are retarded functions and $n_F(\omega)$ is the Fermi function.
At large $U$, and near half-filling, $\langle S_z\rangle \approx 1/2$, and
$\Delta_0 \approx U/2$. For larger dopings and smaller $U$, $\langle S_z\rangle$ is  smaller already at $T=0$.

Fermion-fermion interactions in the ordered SDW state can be cast into
interactions between fermions and magnons. These interactions are described by the effective Hamiltonian~\cite{rep}
 \begin{eqnarray}
\label{inter-hamil}
H_{el-mag}&=&\sum_{\alpha,\beta} \sum_{k,q} \bigl[a^+_{\alpha k} a_{\beta k+q} e^+_q V_{aa}(k,q)\nonumber\\
&+&b^+_{\alpha k} b_{\beta k+q} e^+_q V_{bb}(k,q)
+ a^+_{\alpha k} b_{\beta k+q} e^+_q V_{ab}(k,q)\nonumber\\
&+&b^+_{\alpha k} a_{\beta k+q} e^+_q V_{ba}(k,q)
+ h.c.
 \bigr]\delta_{\alpha,-\beta},
\end{eqnarray}
where operators $a$ and $b$ represent conduction and valence band
fermions, respectively. The vertex functions are given by
 \begin{eqnarray}
\label{vertex}
V_{aa,bb}(k,q)= \frac{U}{\sqrt{2}} \bigl[ \pm (u_k u_{k+q}-v_k v_{k+q})\eta_q \nonumber\\
\qquad +   (u_k v_{k+q}-v_k u_{k+q})\bar{\eta}_q,\nonumber\\
V_{ab,ba}(k,q)= \frac{U}{\sqrt{2}} \bigl[ (u_k v_{k+q}+v_k u_{k+q})\eta_q \nonumber\\
\qquad \mp  (u_k u_{k+q}+v_k v_{k+q})\bar{\eta}_q
 \bigr],
\end{eqnarray}
with
\begin{eqnarray}
  \eta_q = \frac{1}{\sqrt 2}\Bigl( \frac{1-\gamma_q}{1+\gamma_q}\Bigr)^{1/4}, \qquad
  \bar{\eta}_q = \frac{1}{\sqrt 2}\Bigl( \frac{1+\gamma_q}{1-\gamma_q}\Bigr)^{1/4}.
\end{eqnarray}

At the mean-field level, the spectral function $A(k, \omega)$
 has two $\delta-$functional peaks at $\omega = E^{c}_k$ and $\omega = E^v_k$,
 and the density of states has a gap $2\Delta_0 = U$.
Our goal is to analyze how this spectral function gets modified once we
use Eq.~(\ref{inter-hamil}) and compute thermal
  fermionic self-energy and thermal corrections to $\langle S_z\rangle$.

\begin{figure}[h]
\centerline{\includegraphics[width=85mm,angle=0,clip]{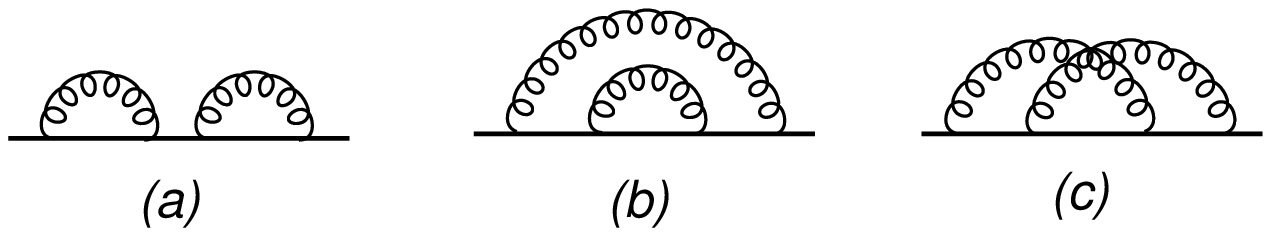}}
\caption{Three equivalent diagrams for two-loop corrections to the Green's function. The second and the third diagrams contribute to fermionic self-energy.}
 \label{fig:1}
\end{figure}

\subsection{One-loop perturbation theory}

 There are several contributions to
 fermionic self-energy $\Sigma (k, \omega)$ to one loop order,
 but the earlier study by Morr and one of us has found~\cite{rep} that the dominant one at the lowest $T$ comes from the
interaction between valence and conduction fermions mediated by the exchange of
low-energy transverse spin-waves (see Ref.~\onlinecite{rep} for details).
 To one-loop order,
 spin-wave mediated interaction gives rise to
\begin{eqnarray}
\label{sigma1cv}
\Sigma_{1}^{v,c}\left(\omega,{\bf k}\right)=\bb \Delta_0^2 G_{c,v}\left(\omega,{\bf k}\right)=\frac{\bb \Delta_0^2}{\omega-\bar{E}_{k}^{cv}},
\end{eqnarray}
where, we remind, $\beta = (T/\pi J) |\log \epsilon|$.
The order parameter  also acquires a correction proportional to $\beta$. To obtain it, one has to substitute the self-energy into the Green's function and
 compute $\langle S_z\rangle$ using Eq.~(\ref{ch_1}), but with the full $G$ instead of $G_0$. This yields
\begin{eqnarray}
\label{delta1}
\langle S_z\rangle= \frac{1}{2}\left(1-{\bb\over 2}\right),
\end{eqnarray}
i.e., $\Delta^2 = \Delta^2_0 (1-\beta + O(\beta^2))$.
The $O(\beta)$ ($|\log \epsilon|$)
correction to $\langle S_z\rangle$ is in agreement with Mermin-Wagner theorem.
However, when we combine self-energies (\ref{sigma1cv}) and
 $G^{v,c}_0$ in which $\Delta_0 = U/2$ is replaced by $\Delta = U \langle S_z\rangle$ and
obtain the new
Green's function, we find that $O(\beta)$ terms  cancel out, i.e.,
 to first order in $\beta$ the fermionic Green's function does not change:
\bea
&&
G(\omega,k, \beta) \approx \frac{1}{2} \left[\frac{1}{\omega - E_c - \Sigma^c_1} + \frac{1}{\omega -E_v - \Sigma^v_1}\right] \nonumber \\
&& = \frac{\bar \omega}{{\bar \omega}^2 - 16t^2 \gamma^2_k - (\Delta^2 + \beta \Delta^2_0)} = G_0 (\omega, k),
\label{ch_2}
\eea
where ${\bar \omega} = \omega + 4t^{\prime} \cos k_x \cos k_y + \mu$.
 This result was obtained in~\cite{rep} and
 was interpreted as an indication that the SDW form of $G(k, \omega)$
  may survive even when $\langle S_z\rangle$ vanishes. However, one-loop result is at best
indicative, and we need to go to higher orders to verify what
 happens with  the fermionic Green's function when $\beta$ increases.

\begin{figure}[t]
\centerline{\includegraphics[width=75mm,angle=0,clip]{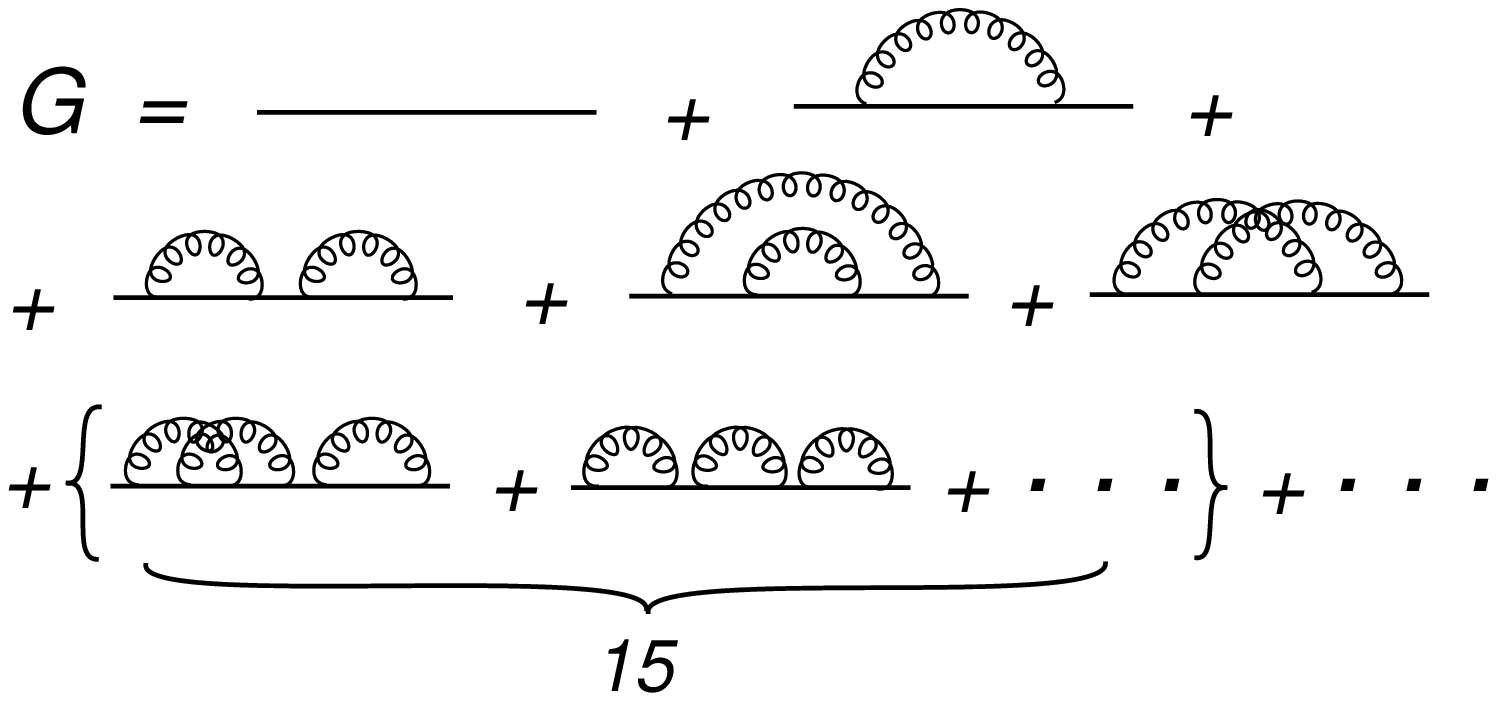}}
\caption{Diagrammatic series for the Green's function. Only thermal
contributions in which a fermion jumps from a valence to a conduction band (and vise versa)  and emits (absorbes) a transverse spin-wave are included.}
 \label{Green-function-chain-2}
\end{figure}

\subsection{Two-loop corrections}
As the next step, we obtain  two-loop formulas for the self-energy and $\langle S_z\rangle$.
The two-loop diagrams for the self-energy are the second and third
 diagrams in Fig.~\ref{fig:1}.
Evaluating them in the same approximation as one-loop diagram,
 we obtain
\begin{eqnarray}
\label{sigma2cv}
\Sigma_2^{c}\left(\omega,{\bf k}\right)=\frac{2\bb^2 \Delta_0^4}{
\left(\omega-E_{k}^{v}\right)\left(\omega-E_{k}^{c}\right)^2},\nonumber\\
\qquad\Sigma_2^{v}\left(\omega,{\bf k}\right)=\frac{2 \bb^2 \Delta_0^4}{\left(\omega-E_{k}^{v}\right)^2\left(\omega- E_{k}^{c}\right)}.
\end{eqnarray}
Substituting these self-energies into the valence and conduction Green's functions together with one-loop diagrams, we obtain after a simple algebra
\begin{eqnarray}
\label{Gv2}
G(\omega,{\bf k})&=& \frac{\bar \omega}{3} \Biggl[\frac{2}{{\bar \omega}^2 - 16t^2 \gamma^2_k - (\Delta^2 + 2 \Delta^2_0 \beta)} \nonumber\\
&+& \frac{1}{{\bar \omega}^2 - 16t^2 \gamma^2_k - (\Delta^2 - \Delta^2_0 \beta)}\Biggr],
\end{eqnarray}
Evaluating $\langle S_z\rangle$ in the same two-loop approximation  we obtain
\be
\langle S_z\rangle = \frac{1}{2} \left(1- \frac{\beta}{2} + \frac{5 \beta^2}{8}\right),
\ee
such that $\Delta^2 = \Delta^2_0 \left(1 - \beta + 3 \beta^2/2 + O(\beta^3)\right)$. Substituting now this $\Delta^2$ into (\ref{Gv2}) we find that, up
to two-loop order,
\begin{eqnarray}
\label{Gv2_a}
G(\omega,{\bf k})= \frac{\bar \omega}{3} \left[\frac{2}{{\bar \omega}^2 - 16t^2 \gamma^2_k - \Delta_1^2} + \frac{1}{{\bar \omega}^2 - 16t^2 \gamma^2_k - \Delta_2^2}\right],\nonumber\\
\end{eqnarray}
where $\Delta_1 = \Delta_0(1 + \beta/2 + 5\beta^2/8)$ and $\Delta_2 = \Delta_0 (1 - \beta + \beta^2/4)$.

We see that the Green's function splits into two components.  Both have SDW form, but the values of $\Delta$ are different -- $\Delta_2$
 decreases with $\beta$ while $\Delta_1$ increases.
 This implies that the peak in the spectral function, originally located
 at ${\bar \omega}^2 = 16 t^2 \gamma^2_k + \Delta^2_0$, splits into two subpeaks -- one shifts to  higher $|{\bar \omega}|$, another to
 smaller $|\bar \omega|$. This is the  new trend, not present in the
 one-loop approximation.

This consideration also shows that to understand what happens when $\beta = O(1)$, one cannot restrict with a few first orders in the loop explansion but rather has to sum up infinite number of terms. This is what we are going to do next.

\subsection{Non-perturbative Green's function}

We list several results which
 can be explicitly  verified by doing loop expansion order by order in $\beta$:
\begin{itemize}
 \item
   the SDW order parameter $\langle S_z\rangle$ is given by the same loop expression as in the mean-field theory, Eq.~(\ref{ch_1}), but with the full $G$ instead of $G_0$
\item
 the renormalized $\langle S_z\rangle$ in turn appears in the Green's function through $G^{v,c}_0$ in which
 $\Delta_0  = U/2$  has to be replaced by $\Delta =U \langle S_z\rangle$
\item
the full Green's function  $G (\omega,{\bf k},\bb)$ is given by
$G (\omega,{\bf k},\bb)= v_k^2G^v(\omega,{\bf k},\bb)+
u_k^2G^c(\omega,{\bf k},\bb)$, where $G^{v,c}$ are the full Green's functions for valence and conduction fermions and $u_k$ and $v_k$ are the same as in
(\ref{twopoles}) but with $\Delta$ instead of $\Delta_0$
\item
 at loop order $n$ of the perturbation theory  there are
 $(2n-1)!!$ equivalent contributions to the full  $G^{c,v}$, each contains
 $\left(\beta G_0^{v}(\omega,{\bf k})G_0^{c}(\omega,{\bf k})\right)^n$  (see  Fig.~\ref{Green-function-chain-2}).
\end{itemize}
\begin{figure}[t]
\centerline{\includegraphics[width=85mm,angle=0,clip]{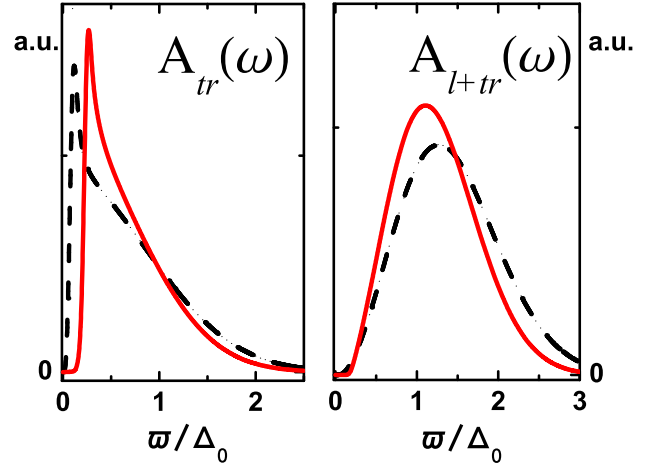}}
\caption{(Color online)
The spectral function $A(\omega, k_{hs}, \beta)$ at a hot spot.
Here and below a.u. stand for arbitary units. Left panel -- $A_{tr}$,
obtained by including only transverse spin waves. Right panel --  $A_{l+tr}$,
 obtained by treating transverse and longitudinal spin excitations on equal footings. Dashed lines -- $\beta = 0.6\beta_{cr}$, solid lines --
 $\beta = 0.8\beta_{cr}$. $A_{tr}$ has a branch cut at  energy which scales with the magnitude of SDW order parameter, while $A_{l+tr}$ has a hump at  energy which roughly remains the same as SDW gap at $T=0$. The actual $A(\omega, k_{hs}, \beta)$ at $\beta < \beta_{cr}$ coincides with $A_{tr}$ at low frequencies, and crosses over to $A_{l+tr}$ at frequencies larger than the gap for longitudinal fluctuations. As a result, the actual spectral function has a peak at a
 low energy and a hump at a higher energy.} \label{hot_spot}
\end{figure}


Because of the last item, it is advantageous to sum up infinite series of
 diagrams for the Green's function rather than for the self-energy.
We have
\begin{eqnarray}
\label{GGv}
G^{v,c}_{tr}(\omega,{\bf k},\bb)&=&G_0^{v,c}(\omega,{\bf k}) +
\sum_{n=1}^{\infty}(2n-1)!!\; G_0^{v,c}(\omega,{\bf k})\nonumber\\
&\times &\Bigl[\bb \Delta^2_0 G_0^{v}(\omega,{\bf k})G_0^{c}(\omega,{\bf k})\Bigr]^n + ...
\end{eqnarray}
where dots stand for non-logarithmic corrections and subindex $tr$ implies that we only considered interaction with transverse spin waves.

Substituting the expressions for $G^{v,c}_0$ and summing up asymptotic series we obtain
\begin{eqnarray}
\label{GF}
&&G^{v,c}_{tr} (\omega,{\bf k},\bb)=\frac{2}{\Delta_0}\left(\frac{\pi}{2\bb}\right)^{1/2}
\frac{{\bar \omega} \mp E_k}{({\bar \omega}^2 - E^2_k)^{1/2}}\\
&\times&\exp\left\{-\frac{{\bar \omega}^2 - E^2_k}{2\Delta_0^2\bb}\right\}~~\left\{i+\text{Erfi}\left[\sqrt{\frac{{\bar \omega}^2 - E^2_k}{2\Delta_0^2\bb}}\right]\right\} + ...,\nonumber
\end{eqnarray}
where $\text{Erfi}(z) = - i \text{Erf}(iz)$ is imaginary error function ($\text{Erfi} (x)$ is real when $x$ is real and imaginary when $x$ is imaginary). Observe that $Im G^{v,c}_{tr}$ vanishes when ${\bar \omega}^2 < E^2_k$.

To one-loop order, Eq.~(\ref{GF}) reduces to $G^{v,c}_0$, but beyond one loop
 the Green's functions $G^{v,c}$ obviously become $\beta$-dependent.
The spectral function $A_{tr} (\omega,{\bf k},\beta)=\pi^{-1}|{\text{Im}}G(\omega-i0,{\bf k},\bb)|$, is readily obtained from (\ref{GF}):
\begin{eqnarray}
\label{spec1}
&&A_{tr} (\omega,{\bf k},\beta)=\frac{2}{\Delta_0}\sqrt{\frac{1}{2\pi\bb}}
e^{\frac{-\left({\bar \omega}^2 - E^2_k\right)}{2\Delta_0^2\bb}}\nonumber \\
&&\times\frac{u_k^2 \vert {\bar \omega} + E_k\vert +v_k^2  \vert{\bar \omega}-E_k\vert}{({\bar \omega}^2 -E^2_k)^{1/2}}
\theta\Bigl[{\bar \omega}^2 -E^2_k\Bigr],
\end{eqnarray}
where $\theta (x) =1$ for $x >0$ and zero otherwise.

Substituting  $G^{v,c}$ from (\ref{GF}) into the expression for~~~
 $\langle S_z\rangle$ we obtain how the
 SDW order parameter evolves with $\beta$:
\begin{eqnarray}
\label{GapG}
 \langle S_z\rangle  &=& \frac{1}{2}\int\frac{d^2k}{(2\pi)^2} \;
\sqrt{\frac{2}{\pi\bb}} \frac{\Delta}{\Delta_0}\int_{-\infty}^{\infty}d{\bar \omega}
\frac{\exp\left\{-\frac{{\bar \omega}^2 -E_k^2}{2\Delta_0^2\bb}\right\}}
{\sqrt{{\bar \omega}^2 -E_k^2}}\nonumber\\
& \times &n_F(\bar{\omega}-\mu-4t^{\prime}\cos k_x\cos k_y)
\theta\Bigl[{\bar \omega}^2 - E^2_k\Bigr],
\end{eqnarray}
The remaining unknown parameter $\mu$ is fixed by the condition on the number of particles in the SDW state~\cite{Lutt}:
\begin{eqnarray}
\label{mu1}
&&\frac{(1-x)}{2}=  \int\frac{d^2k d\omega}{(2\pi)^2}
A_{tr}(\omega,{\bf k}, \beta) n_F(\omega).
\end{eqnarray}

\begin{figure}[t]
\centerline{\includegraphics[width=85mm,angle=0,clip]{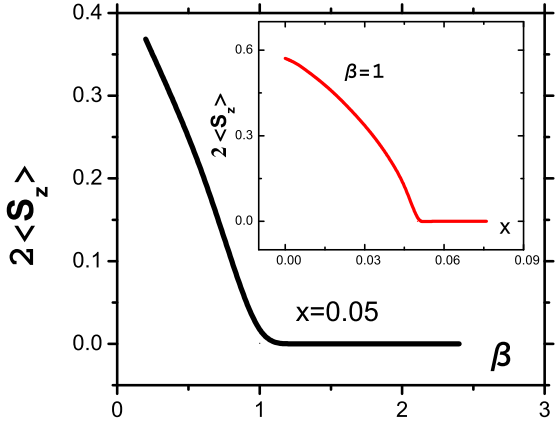}}
\caption{(Color online) The SDW order parameter $\langle S_z\rangle$ vs $\beta$  at a given
 $x=0.05$. For simplicity, we set $Z_\beta$ to be a constant ($=0.2$).
 Inset:  $\langle S_z\rangle$  vs $x$ at a given $\beta =1$.} \label{Gaps}
\end{figure}

These coupled equations were solved numerically.
The dependence of $\langle S_z\rangle$ on $\beta$ or doping $x$ is quite as expected: $\langle S_z\rangle$ monotonically decreases as $\beta$ or $x$ increase, and vanishes at some particular $\beta_{cr}$ and $x_{cr}$ (see Fig.~\ref{Gaps}).
The spectral function $A_{tr} (\omega, {\bf k}, \beta)$
 has  sharp $\delta-$functional
peaks at $\beta \rightarrow 0$, at ${\bar \omega}  = \pm E_k$ ($\omega = E_k^{v,c}$). At finite $\beta$, quasiparticle peaks
transform into branch cuts at $|{\bar \omega}| = E_k$,
 with the width of order $\beta \propto T$,
and  the spectral weight extends to larger frequencies. Near the branch cut
$A_{tr} (\omega, {\bf k}, \beta)$
diverges as   $1/\sqrt{x}$. We plot $A_{tr}  (\omega,{\bf k},\beta)$ at a hot spot ${\bf k}_{hs} = (k_x, \pi-k_x)$ in Fig.~\ref{hot_spot}.

\subsection{Further modifications of the spectral function}

 On a more careful look, we found that the spectral function given by
Eq.~(\ref{spec1}) have to be further modified by two reasons.
First, in the calculations above we only included the self-energy due to exchange of transverse spin waves, and neglected the self-energy due to exchange of longitudinal spin fluctuations. This is justified at small $\beta$, when
  longitudinal fluctuations are gapped, but when $\beta \approx \beta_{cr}$, longitudinal fluctuations are nearly gapless and are as important as transverse ones. The limiting case when transverse and longitudinal spin propagators are identical can be studied within the same approximation as before, the only difference is that combinatoric factors are now $(2n+1)!!/2^n$ (Ref.~\onlinecite{JS}).  As a result, the spectral function becomes
\begin{eqnarray}
\label{spec1_1}
&&A_{l+tr}(\omega,{\bf k},\beta)=\frac{1}{\Delta^3_0} {\frac{4}{\sqrt{\pi\bb^3}}}
 \exp \Biggl\{\frac{-({\bar \omega}^2 - E^2_k)}{\Delta_0^2\bb}\Biggr\}
\nonumber \\
&&\times\frac{u_k^2 \vert {\bar \omega} +E_k\vert +v_k^2  \vert{\bar \omega}-E_k\vert}{({\bar \omega}^2 - E^2_k)^{-1/2}}
\theta\Bigl[{\bar \omega}^2 - E^2_k\Bigr],
\end{eqnarray}
where subindex $l+tr$ implies that this is a contribution from both longitudinal and transverse spin excitations.  One can easily make sure that $A_{l+tr}$ is also fully incoherent at $\beta >0$, but, in distinction to $A_{tr}$, it
 vanishes at ${\bar \omega} = \pm E_k$ and has a hump at a frequency which remains
 of order $\Delta_0$ for all $\beta < \beta_{cr}$
 We plot  $A_{l+tr}  (\omega,{\bf k},\beta)$ at a hot spot in Fig.~\ref{hot_spot}. The function $A_{l+tr}(\omega,{\bf k}_{hs},\beta)$  becomes particularly simple at $\beta \geq \beta_{cr}$:
\begin{eqnarray}
\label{specHS}
A_{l+tr}(\omega,{\bf k}_{hs})=\frac{\omega^2}{\Delta_0}\frac{4}
{\sqrt{\pi \bb^{3}}}\exp
\Biggl\{-\frac{\omega^2}{\bb \Delta_0^2}\Biggr\},
\end{eqnarray}

For a generic $\beta \leq \beta_{cr}$, Eq.~(\ref{spec1_1}) is the correct result at high energies, larger than the longitudinal gap, and  Eq.~(\ref{spec1}) is the correct result at smaller energies.  As $\beta$ approaches $\beta_{cr}$ the range of applicability of $A_{tr} (\omega,{\bf k}_{hs})$ shrinks. The full formula cannot be obtained within eikonal approximation, but it is clear that the
 actual $A (\omega,{\bf k},\beta)$ contains both, the branch cut, $1/\sqrt{x}$ singularity near $\omega = E^{v,c}_k$ and the hump at a frequency of order $\Delta_0$, where the quasiparticle peaks were located at $\beta =0$.
 To simplify the computational procedure, in Fig.~\ref{HS_COMBINED} below we use for the actual $A (\omega, k_{hs} ,\beta)$ the function
 $A_{tr} (\omega, k_{hs},\beta)$ up to a frequency where it crosses with
$A_{l+tr}$, and use the function $A_{l+tr} (\omega, k_{hs}, \beta)$ at larger frequencies.

Second,  Eqs.~(\ref{spec1}) and (\ref{spec1_1})  show that
 fermionic coherence is lost immediately when $\beta$ becomes non-zero (the pole transforms into a branch cut).
Meanwhile, from physics perspective, as long as the system has a SDW order, a
 Fermi liquid behavior near the pocketed FS should be preserved, i.e., a quasiparticle peak with $T^2 \log T$ width should survive, albeit with a reduced magnitude. The reason it was lost in the calculations above is because we completely neglected regular classical and quantum corrections to the Green's function (dots in Eq.~(\ref{GGv})).
We verified that, when these terms are included,
only a part of  $G^{v,c}_0$ gets involved in the renormalizations by
 series of $\beta^n$ corrections, the other stays intact. This implies that the
the actual $A^{v,c}_{full} = Z_\beta A^{v,c}_0 + (1-Z_\beta) A^{v,c}$.
The residue $Z_\beta$ is some number $0<Z_\beta <1$ at
 $\beta =0$, where anyway $A^{v,c} = A^{v,c}_0$, it
decreases as $\beta$ increases, and vanishes at $\beta = \beta_{cr}$.
For definiteness, we used $Z_\beta=0.2$ in the panel $0< \beta < \beta_{cr}$
 in Figs. \ref{spectral-nodal} and \ref{HS_COMBINED}.

These two additions also affect the formula for $\langle S_z\rangle$ and the equation for $\mu$, which become

\begin{widetext}
\begin{eqnarray}
\label{GapG_1}
 \langle S_z\rangle  &=& Z_{\bb} \int\frac{d^2k}{(2\pi)^2} \frac{\Delta}{E_k} \left[n_F (E^v_k) - n_F (E^c_k)\right] \\
&+& (1-Z_{\bb})
\frac{1}{2}\int\frac{d^2k}{(2\pi)^2} \;
\sqrt{\frac{2}{\pi\bb}} \frac{\Delta}{\Delta_0}\int_{-\infty}^{\infty}d{\bar \omega}
\frac{\exp\left\{-\frac{{\bar \omega}^2 -E_k^2}{2\Delta_0^2\bb}\right\}}
{\sqrt{{\bar \omega}^2 -E_k^2}}n_F(\bar{\omega}-\mu-4t^{\prime}\cos k_x\cos k_y)
\theta\Bigl[{\bar \omega}^2 - E^2_k\Bigr],
\nonumber
\end{eqnarray}
and
\begin{eqnarray}
\label{mu1_1}
\frac{(1-x)}{2}=\int_{-\infty}^{\infty}d\omega \int \frac{d^2k}{(2\pi)^2} n_F(\omega)
\times\left\{\frac{Z_{\bb}}{\pi}
\left[v^2_k\delta(\omega-E_k^v)+u^2_k\delta(\omega-E_k^c)\right]+
(1-Z_{\bb})A(\omega,{\bf k})\right\}.
\end{eqnarray}

\begin{center}
\begin{figure}[t]
\centerline{\includegraphics[width=175mm,angle=0,clip]{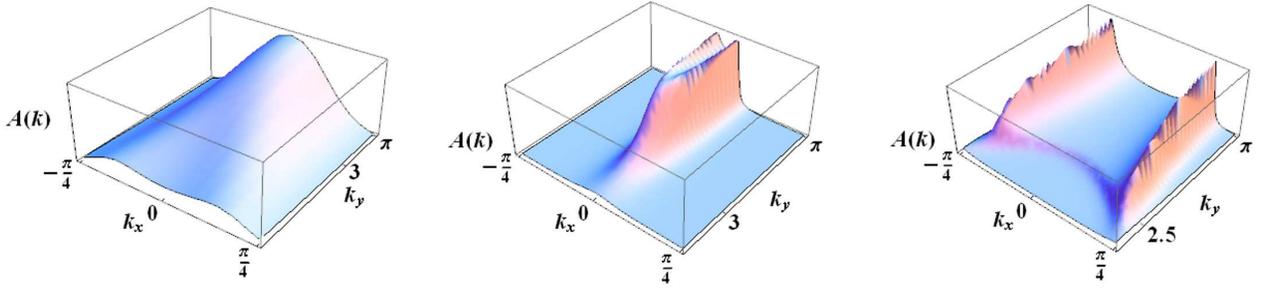}}
\caption{(Color online) The appearance and evolution of the electron
 pocket near $(k_x,k_y)=(0,\pi)$ and symmetry related points.
 Electron pocket appears as a single point at $(0,\pi)$ once $\bb$ increases and reaches a critical value, and evolves with increasing $\bb$.
 From left to right: $\bb/\bb_{cr}=0.8;\;0.85;\;0.95$. We set $x=0.02$. } \label{3D}
\end{figure}
\begin{figure}[h]
\centerline{\includegraphics[width=145mm,angle=0,clip]{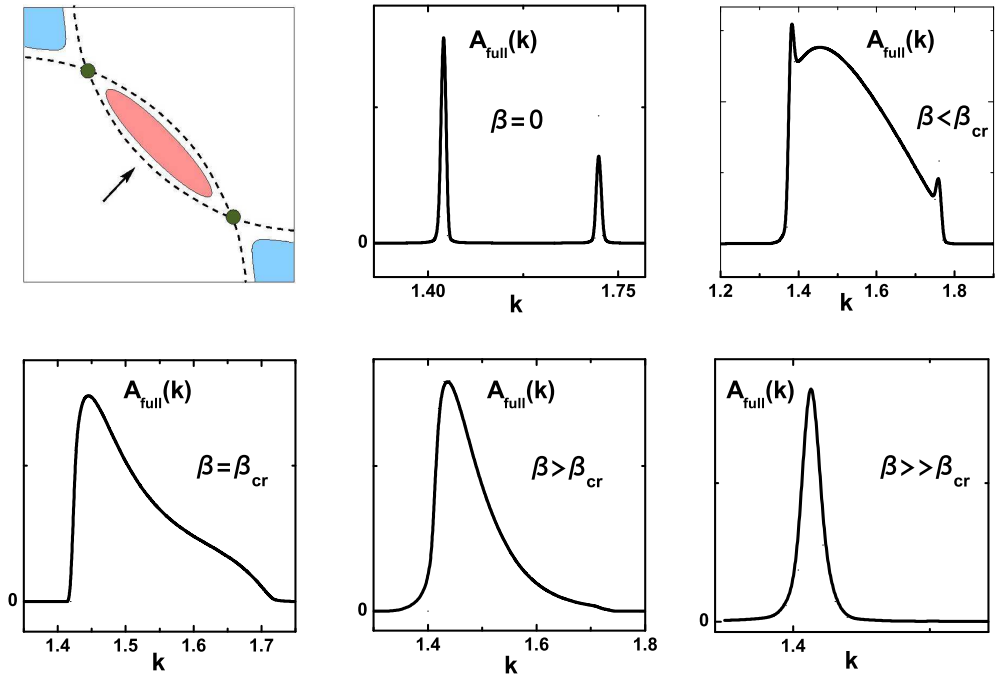}}
\caption{(Color online) The spectral function, $A_{full}
(\omega,{\bf k},\beta)$ at  $\omega=0$ along the diagonal (nodal) direction in the Brillouin zone. We used $Z=0.2$ for the top right panel.
 Observe that the systems retains a memory about a ``shadow'' FS  when SDW order disappears at $\beta = \beta_{cr}$ (the spectral weight is non-zero in the whole region between the original and the shadow FSs).
 We set $x=0.05$, $t = 0.32 \Delta_0$, $t^{\prime} =-0.2t$.
 For these parameters, $\beta_{cr} \approx 1$.} \label{spectral-nodal}
\end{figure}
\begin{figure}[h]
\centerline{\includegraphics[width=145mm,angle=0,clip]{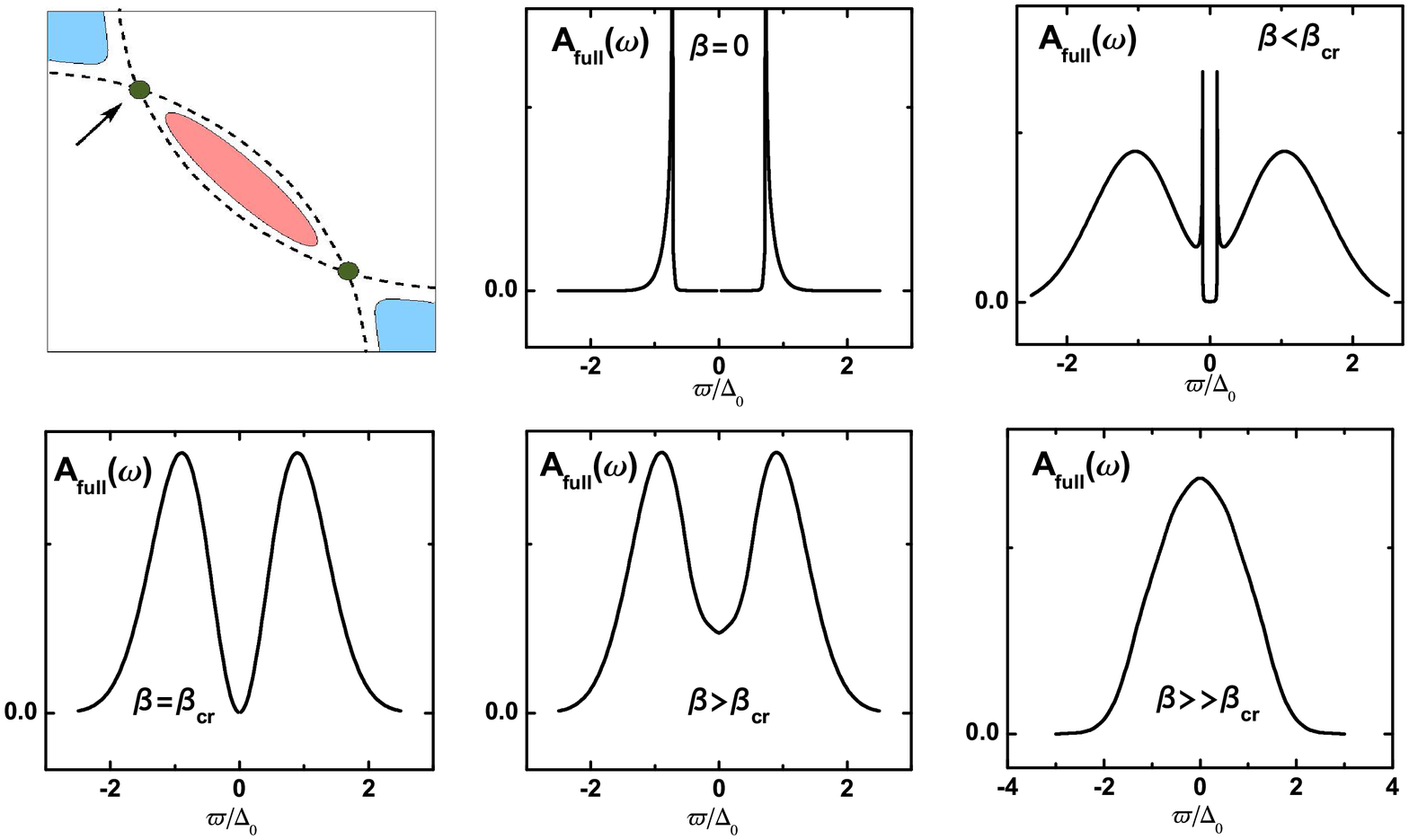}}
\caption{(Color online)
The spectral function  $A_{full}(\omega,{\bf k}_{hs},\beta)$ at a hot spot, at various $\beta$. The frequency is in units ${\bar{\omega}}/\Delta_0$, where ${\bar \omega} = \omega + \mu - 4 t^{\prime} \cos^2k_x$.The parameters are the same as in Fig.~\protect\ref{spectral-nodal}.}
\label{HS_COMBINED}
\end{figure}
\end{center}
\end{widetext}

\section{The results}

We solved Eqs.~(\ref{GapG_1}) and (\ref{mu1_1}) numerically and
plot the dependence of the order order parameter $\langle S_z\rangle$ on
$\beta$ and $x$ in Fig.\ref{Gaps}. We used several phenomenological forms of $Z_\beta$ in which $Z_{\beta_{cr}} =0$, but found that the functional forms of the spectral functions do not depend on $Z_\beta$ in any substantial way. For
 simplicity, and because below we only present the results for a few $\beta$ in the SDW state, we set $Z_\beta = 0.2$, independent on the actual $\beta$.
 Also,  all cases (even when $Z_\beta =0$)  $\langle S_z\rangle$ monotonically decreases when either $x$ or $\beta$ increase, and vanishes at some $x_{cr}$ and
$\beta_{cr}$.

 In Fig. \ref{3D} we show the evolution of the electron Fermi surface
 near $(0,\pi)$ and symmetry-related points. The electron pocket is absent at small $\beta$, when SDW order is strong,  but appears when $\beta$ exceeds some critical value. The electron pocket evolves with increasing $\beta$ and eventually disappers when the system looses long-range SDW order. This behavior is qualitatively consistent with the mean-field SDW picture.

\begin{figure}[h]
\centerline{\includegraphics[width=95mm,angle=0,clip]{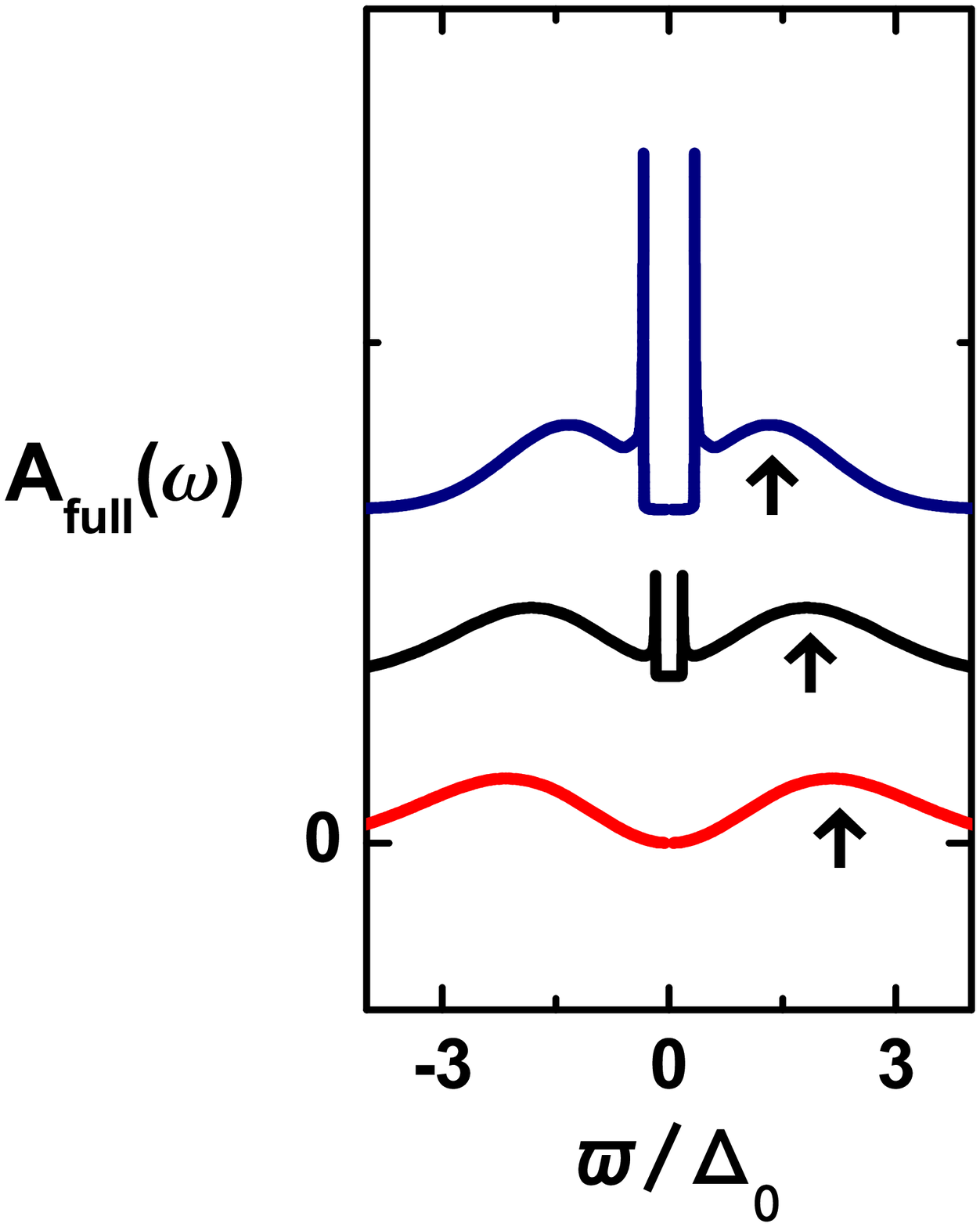}}
\caption{(Color online) The spectral function
$A_{full}(\omega,{\bf k}_{hs},\beta)$ at a hot spot, at various $0<\beta\leq\bb_{cr}$. The lowest plot corresponds to $\bb=\bb_{cr}$,  values of $\bb$
decrease from bottom up. The frequency is in units ${\bar \omega}/\Delta_0$. The parameters are the same as in Fig.~\protect\ref{spectral-nodal}. As $\beta$ decreases, the peak and the hump come closer to each other.} \label{Gaps_1}
\end{figure}
The results for  $A_{full} (\omega, k, \beta)$ are presented in Figs.  \ref{spectral-nodal}, \ref{HS_COMBINED}, and \ref{Gaps_1}.
  In  Fig.~\ref{spectral-nodal} we plot the full spectral function at zero frequency along the diagonal direction in the Brillouin zone.
In the two limits, $\beta=0$ and $\beta \gg \beta_{cr}$, the system possesses
 sharp quasiparticles -- in the first case at the two sides of the hole pocket, in the second case at the large,  Luttinger FS.
In between, the spectral function evolves, as $\beta$ increases,
 from a well-pronounced two-peak structure to a completely incoherent structure
 at $\beta = \beta_{cr}$, in which the spectral weight at $\omega =0$ is spreaded between the original and the ``shadow'' FSs (in reality,
$A_{full} (0, k, \beta)$ spreads outside of this range, but to find these tails of the spectral function one has to go beyond the accuracy of our calculations).
This result implies that the system does retain some memory about SDW pockets
even when $\beta = \beta_{cr}$ and $\Delta =0$.
 When $\beta$ becomes larger than $\beta_{cr}$ and SDW order disappears, the region where $A_{full} (0, k, \beta) \neq 0$ progressively shrinks with increasing $\beta$ towards  a single quasiparticle peak.  We emphasize that
 this evolution of $A_{full} (0,k, \beta)$ with $\beta$
 is very different from that in the
 mean-field SDW theory, where the spectral function in the SDW state
has  two peaks at the two sides of the hole pocket, and the ``shadow'' peak just disappears when $\Delta$ vanishes.

In Fig.~\ref{HS_COMBINED} we show the frequency dependence of $A_{full} (\omega, k_{hs}, \beta)$ at a hot spot for a wide range of $\beta$. In the two limits, the behavior is again coherent: there are quasiparticle peaks at
${\bar \omega} = \omega + \mu - 4 t^{\prime} \cos^2k_x = \pm \Delta_0$
 deep in the SDW phase, and the peak centered at  $\bar \omega = \omega= 0$
 deep in the normal phase. In between, the spectral function is again predominantly incoherent. Specifically, as $\beta$ increases towards $\beta_{cr}$,
 the quasiparticle peak splits into the peak at  ${\bar \omega} = \Delta$, and the hump at a larger frequency.
The frequency where the peak is located  shifts
 downwards with decreasing $\beta$, while the hump  remains roughly
 at around $\Delta_0$, and, on a more careful look,
 shifts towards somewhat larger frequency. We show this behavior in more detail in Fig.\ref{Gaps_1}, where we
 plot the spectral function for a range of $\beta < \beta_{cr}$.
The peak in the spectral function
 is the property of  $A_{tr}$ and $A_0$, and the hump is the property of $A_{l+tr}$. We emphasize that the peak and the hump do co-exist in the SDW phase, the first
 describes  coherent  low-energy excitations, the second describes fully incoherent high-energy excitations.  Once SDW order disappears at $\beta_{cr}$, coherent excitations also disappear, but the hump remains and disappears only at much larger $\beta$.  It is quite natural to identify the hump at $\beta > \beta_{cr}$
 with the pseudogap, and low-energy coherent excitations existing
at $\beta < \beta_{cr}$ with the building blocks for magneto-oscillations.
 The implication of this result is that  the $200 meV$ pseudogap observed in
 ARPES in zero field~\cite{arpes_pgap}
 {\it is not} an obstacle for observing magneto-oscillations once the system becomes SDW-ordered.

\subsection{Density of states}

 The density of states (DOS) in the SDW phase, $N(\omega,\beta)=\int (d^2k/4\pi^2)
 A(\omega,{\bf k}, \beta)$, is obtained from Eqs.~(\ref{twopoles}), (\ref{spec1}), and (\ref{spec1_1}). The expressions for $N(\omega)$ at finite $\beta$
are rather complex for $t^{\prime} \neq 0$
 but are simplified for $t^{\prime}=0$ which we assume to hold in this subsection.
A finite $t'$ affects the behavior of the DOS at the smallest frequencies, but
 not at frequencies $\omega + \mu \geq  \Delta$,
 which we are chiefly interested in.

  For free fermions, we then have
\bea
&&N_0(\omega,\beta ) = \\
&&\frac{\tw}{\pi^2} \int\int_{-1}^1 \frac{du dv}{\sqrt{1-u^2} \sqrt{1-v^2}} \delta\left(\tw^2 -\Delta^2 -4t^2 (u+v)^2\right) \nonumber \\
&&=\frac{1}{2\pi^2 t} \frac{{\bar \omega}}{\sqrt{{\bar \omega}^2 - \Delta^2}} K\left[\sqrt{1-\frac{{\bar \omega}^2 - \Delta^2}{16t^2}}
\right],\nonumber
\label{ch_3}
\eea
where $K(x)$ is the complete elliptic integral of the first kind (defined as in Ref.~\onlinecite{gradstein}), and for $t^{\prime}=0$, $\bar \omega = \omega + \mu$.
The DOS vanishes at $|\tw| <\Delta$, diverges as $1/\sqrt{x}$ at  $|\tw| =
\Delta +0$, monotonically decreases at larger frequencies, and discontinuously drops to zero at the bandwidth, when $|\tw| = \sqrt{\Delta^2 + 16 t^2}$.

The DOS $N_{tr} (\omega, \beta)$ obtained using the spectral function
$A_{tr}$ from (\ref{spec1}) reduces to a 1D integral
\begin{eqnarray}
\!\!\!\!&&N_{tr} (\omega, \beta) =  \frac{{\bar \omega}}{2 \pi^{5/2} t} \int_0^{\sqrt{\frac{\tw^2-\Delta^2}{2\beta \Delta^2_0}}} \frac{dz e^{-z^2}}{\sqrt{\tw^2 -\Delta^2 -2\beta \Delta^2_0 z^2}} \nonumber\\
\!\!\!\!&&\times K\left[\sqrt{1-\frac{\tw^2-\Delta^2-2\beta\Delta^2_0 z^2}{16t^2}}\right],
\label{ch_4}
\end{eqnarray}
for $0<\tw^2-\Delta^2 < 16t^2$, and
\begin{eqnarray}
\!\!\!\!&&N_{tr} (\omega, \beta) =  \frac{{\bar \omega}}{2 \pi^{5/2} t} \int_{\sqrt{\frac{\tw^2-\Delta^2-16t^2}{2\beta \Delta^2_0}}}^{\sqrt{\frac{\tw^2-\Delta^2}{2\beta \Delta^2_0}}} \frac{dz e^{-z^2}}{\sqrt{\tw^2 -\Delta^2 -2\beta \Delta^2_0 z^2}}\nonumber\\
\!\!\!\!&&\times K\left[\sqrt{1-\frac{\tw^2-\Delta^2-2\beta\Delta^2_0 z^2}{16t^2}}\right],
\label{ch_5}
\end{eqnarray}
for  $\tw^2-\Delta^2 > 16t^2$.
 [The trick how to integrate over 2D momenta $k_x$ and $k_y$ in $\int d^2k/(4\pi^2) A_{tr} (\omega, {\bf k}, \beta)$
  is to introduce $u =\cos k_x$ and $v = \cos k_y$ as new variables, use the identity
\be
\frac{e^{-\frac{\tw^2-E^2_k}{2\beta \Delta^2_0}}}{\sqrt{\tw^2-E^2_k}}
2 \sqrt{2\beta \Delta^2_0}  = \int_0^\infty dz e^{-z^2} \delta ({\bar \omega}^2-E^2_k -2\beta \Delta^2_0 z^2),
\label{ch_3_1}
\ee
and use $\delta-$function to perform the momentum integration].
The density of states $N_{tr}$ vanishes at $|\tw| <\Delta$,
diverges logarithmically at $|{\bar \omega}| = \Delta +0$,
 and monotonically decreases at larger frequencies.  There is no threshold at the bandwidth as $A_{tr} (\omega, {\bf k}, \beta)$ is nonzero everywhere at $|\tw| > \Delta$,  but, indeed, at frequencies larger than the bandwidth
 our approximation eventually breaks down.

\begin{figure}[t]
\centerline{\includegraphics[width=90mm,angle=0,clip]{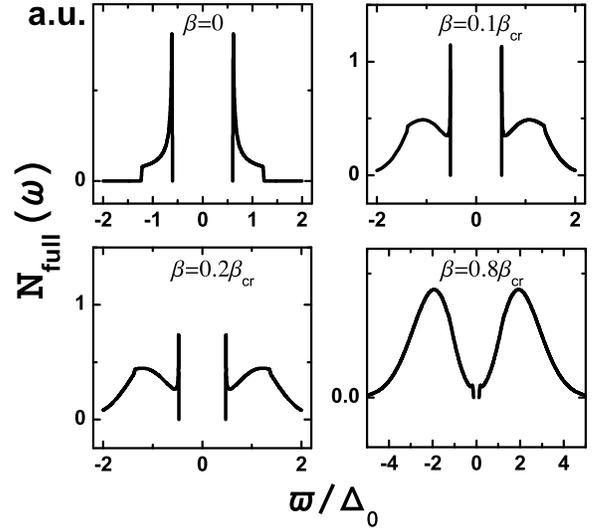}}
\caption{Density of states vs  $\bar\omega/\Delta_0$ for
different values of $\bb$. We set  $t^{\prime}=0$ and $t = 0.32\Delta_0$.} \label{DOS}
\end{figure}

For the spectral function $A_{l+tr}$ from (\ref{spec1_1}), we obtain,
using the same trick,
\begin{eqnarray}
\!\!\!\!\!&&N_{l+tr} (\omega, \beta) = \frac{{\bar \omega}}{\pi^{5/2} t} \int_0^{\sqrt{\frac{\tw^2-\Delta^2}{\beta \Delta^2_0}}} \frac{dz z^2 e^{-z^2}}{\sqrt{\tw^2 -\Delta^2 -\beta \Delta^2_0 z^2}} \nonumber\\
\!\!\!\!\!&&\times K\left[\sqrt{1-\frac{\tw^2-\Delta^2-\beta\Delta^2_0 z^2}{16t^2}}\right],
\label{ch_4_1}
\end{eqnarray}
for $0<\tw^2-\Delta^2 < 16t^2$, and
\begin{eqnarray}
\!\!\!\!\!&&N_{l+tr} (\omega, \beta) = \frac{{\bar \omega}}{\pi^{5/2} t} \int_{\sqrt{\frac{\tw^2-\Delta^2-16t^2}{\beta \Delta^2_0}}}^{\sqrt{\frac{\tw^2-\Delta^2}{\beta \Delta^2_0}}} \frac{dz z^2 e^{-z^2}}{\sqrt{\tw^2 -\Delta^2 -\beta \Delta^2_0 z^2}} \nonumber\\
\!\!\!\!\!&&\times K\left[\sqrt{1-\frac{\tw^2-\Delta^2-\beta\Delta^2_0 z^2}{16t^2}}\right],
\label{ch_5_1}
\end{eqnarray}
for $\tw^2-\Delta^2 > 16t^2$
This spectral function is continuous at  $|\tw| = \Delta$ and has a broad maximum at frequencies comparable to $\Delta_0$.

This behavior is very similar to the one for the spectral function. Again,
the transverse-only contribution $N_{tr}$ has a peak at the frequency
$\tw =  \Delta$, which scales with the order parameter of the SDW phase, while
$N_{l+tr}$ which treats contributions from transverse and longitudinal fluctuations on equal footings (as if longitudinal excitations were massless) has no features at $\Delta$, but has a broad maximum at a frequency which remains comparable to $\Delta_0$.  Just like the spectral function, the actual DOS $N (\omega, \beta)$ interpolates between these two terms. At frequencies smaller than the gap for longitudinal spin excitations, $N(\omega, \beta) \approx N_{tr} (\omega, \beta)$, while at larger frequencies $N(\omega, \beta)$ gradually approaches $N_{l+tr} (\omega, \beta)$.

We plot the DOS in Fig.\ref{DOS}. We again set  $N(\omega, \beta) = N_{tr} (\omega, \beta)$ at frequencies smaller than the crossing point between $N_{tr} (\omega, \beta)$ and
$N_{l+tr} (\omega, \beta)$, and set $N (\omega, \beta) =
N_{l+tr} (\omega,\beta)$ at higher frequencies.
And we again assumed that the full DOS $N_{full} (\omega,\beta)$ is
the sum of  the incoherent $N (\omega, \beta)$ with the factor $1-Z_\beta$ and the coherent, mean-field $N_0 (\omega, \beta)$ with the factor $Z_\beta$.

The plots clearly show the same trends in the DOS as we just discussed.
 At small $\beta$, there is a sharp gap $2\Delta_0 \approx
U$ between valence and conduction bands. As $\beta$ increases, sharp gap decreases, the spectral weight extends to higher frequencies, and the DOS develops a hump at an energy comparable to $\Delta_0$ (i.e., the distance between the humps remains $U$). The peak in the DOS at the boundary of the sharp gap gets smaller as $\beta$ increases and the sharp gap decreases. At $\beta = \beta_{cr}$ the gap and the peak disappear, and the spectral function only possesses a hump. As $\beta$ increases even further, the DOS at $\tw =0$ increases (and also $\mu$
 gets reduced such that $\tw$ comes closer to $\omega$), and eventually the DOS  at low frequencies recovers weakly-frequency-dependent form of a metal with a large FS.

This physics is somewhat spoiled in the plots  of
$N_{full} (\omega, \beta)$ by the change of the behavior of $N_{tr+l} (\omega, \beta)$ at the bandwidth (at ${\bar \omega} = \sqrt{\Delta^2 + 16 t^2}$).
 This change of behavior is seen in Fig.~\ref{DOS}
as the discontinuity in the frequency derivative of $N_{full} (\omega, \beta)$.
 For large $\beta$ the hump in the DOS is predominantly the effect of the bandwidth. However, at at intermediate $\beta$, the hump is located at a frequency below the bandwidth, as is clearly visible in the top right panel of Fig.~\ref{DOS} ($\beta=0.1\beta_{cr}$), and is therefore due to the physics that we described above.

\subsection{Optical conductivity}

\begin{figure}[t]
\centerline{\includegraphics[width=85mm,angle=0,clip]{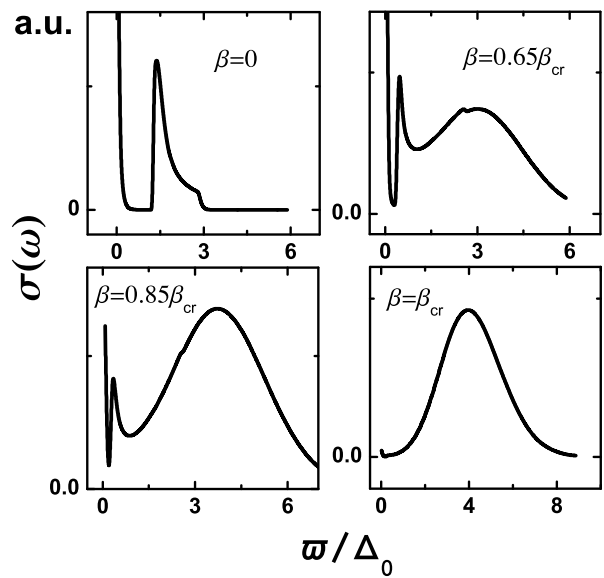}}
\caption{Optical conductivity, $\sigma(\omega)$,
 for different $\beta$ in the SDW phase at $x=0.02$.
For simplicity, we used $Z_\beta = 0.12$ in all plots.
At $\beta=0$ $\sigma (\omega)$ has a true gap and discontinuity at the onset of scattering between conduction and valence bands. The onset frequency is at $\bar\omega$ slightly smaller than  $2\Delta_0 =U$ because $x$ is non-zero and the actual gap $\Delta < \Delta_0$.
 At finite $\beta$ the discontinuity splits into a hump which slowly moves to a higher frequency and a peak  which moves to a smaller energy. We verified that the peak position scales with $ \langle S_z\rangle$.}
 \label{Cond}
\end{figure}

The peak/hump structure also shows up in the optical conductivity.
We computed the conductivity by standard means: by  convoluting two full spectral functions $A_{full}$ using Kubo formula:
\begin{eqnarray}
\label{kubo}
&&\sigma(\omega) =\pi e^2 \int\frac{d^2k}{(2\pi)^2} v^2_k
\int d\Omega \frac{n_F(\Omega)-n_F(\Omega+\omega)}{\omega} \nonumber \\
&& A_{full}(\Omega,{\bf k},\beta)
A_{full}(\Omega + \omega,{\bf k},\beta),
\end{eqnarray}
 where $v_k$ is fermionic velocity.
We used the same computational procedure as before: combined mean-field contribution to $A$, with the prefactor $Z_\beta$, and incoherent part of $A$ with the prefactor $1-Z_\beta$, and used $A_{tr}$ for the incoherent part at small frequencies and $A_{l+tr}$ at high frequencies.  The results are shown in Fig.~\ref{Cond}. We again set $t^{\prime} =0$ to simplify the calculations.
At small $\beta$, the conductivity almost vanishes up to ${\bar \omega} = 2 \Delta_0 =U$. When $\beta$ gets larger, the discontinuity at $U$ splits into a hump which slowly shifts to a larger energy and a peak which scales with $\Delta$
 and shifts to a smaller frequency as  SDW order gets weaker. In addition, there also appears a Drude component at the smallest frequencies.

The hump at large $\beta \geq \beta_{cr}$  is predominantly the effect of the bandwidth. It appears quite ``sharp'' in the last panel in Fig.  \ref{Cond}, but this is the consequence of our approximation in which we neglected all regular self-energy terms. When these terms are included, the hump should definitely get broader.

 This behavior of $\sigma (\omega)$
 is quite consistent with the observed evolution of $\sigma (\omega)$ in the SDW phase of electron-doped $Nd_{2-x}Ce_xCuO_4$, where the region of SDW-ordered phase extends over a substantial doping range, up to $x \sim 0.15$.
 Onose et al~\cite{Onose04}
 and others~\cite{millis_opt} have found that conductivity does have a peak/hump structure at finite $x$, the hump shifts towards a higher frequency with increasing $x$  (from a charge-transfer gap of $1.7eV$ at $x=0$ to over $2eV$ at $x = 0.1$), while the peak shifts downwards as $x$ increases, and was argued~\cite{millis_opt} to scale with the Neel temperature $T_N$. These two features are reproduced in our theory.

\section{Conclusions}

To summarize, in this paper we obtained fermionic spectral function, the density of states, and optical conductivity in the SDW phase of the cuprates at small but finite $T$. We adopted  non-perturbative approach and summed up infinite series of thermal self-energy terms,
keeping at each order nearly-divergent $T/J |\log \epsilon|$ terms, where $\epsilon$ is a deviation from a pure 2D, and neglecting regular $T/J$ corrections. We found that, as SDW order decreases,  the spectral function in the antinodal region acquires peak/hump structure: the peak  position scales with the
SDW order parameter, while the incoherent hump remains roughly at the same scale as at $T=0$, when SDW order is the strongest. We identified the hump with the pseudogap observed in ARPES experiments and identified coherent, Fermi liquid excitations at low energies as  building blocks for
 magneto-oscillations in an applied field. The same peak/hump structure appears in the DOS and in the optical conductivity. The gap in the DOS
scales with the SDW order parameter and disappears when SDW order vanishes,
 however the DOS also develops a hump at an energy which remains close to $U/2$, no matter what is value of the SDW order parameter.  Optical conductivity
 at finite $\beta$ has a Drude peak at the lowest frequencies, a
 peak, which again scales with
the SDW order parameter and moves to smaller frequencies as $\beta$ increases, and a hump which remains roughly at
 $\omega+\mu \sim  U$ and slightly shifts to higher frequencies as SDW order gets weaker.

A more generic result of our study is the phase diagram for the cuprates shown in Fig.~\ref{phase-space}. A similar phase diagram has been proposed in
Ref.~\onlinecite{Sachdev2}. At large enough hole doping and small enough temperatures, thermal fluctuations are weak and no SDW precursors appear. In this region, FS is large, and the physics is governed by the interaction between fermions and Landau-overdamped spin fluctuations. This interaction
 gives rise to a fermionic self-energy which is Fermi liquid-like at the lowest energies, has a non-Fermi liquid form, $(i\omega)^{a}$, $a <1$ at high energies,
and displays a marginal Fermi liquid behavior in the cross-over region~\cite{acs}.  The same interaction with overdamped spin fluctuations
 gives rise to a $d-$wave pairing instability.~\cite{acf,scal,pines}
 The onset temperature for the $d-$wave pairing  increases as $x$ decreases
 and approaches the universal scale of around $T_p \sim 0.02 v_F/a$, where $v_F \sim 1 eV *a$ is the Fermi energy (Ref.~\onlinecite{mike_artem}).
 The actual superconducting $T_c$ is lower
 due to fluctuations of the pairing gap. In this regime, the thermal evolution of the FS is entirely due to thermal effects associated
 $s-$wave fermionic damping induced by scattering on thermal bosons.~\cite{millis_franz,arcs} In particular, at a finite $T$,  the spectral function is peaked at zero frequency in some range of $k$ around the nodal direction, despite that the pairing gap itself has $\cos 2\phi$ form.

At larger $T$ and smaller $x>0$, thermal fluctuations get stronger and give rise to SDW precursors. These does not imply that the FS actually becomes pocket-like, but the spectral function in the antinodal region develops a hump at a finite frequency, and the low energy spectral weight
 progressively fills in the area between the actual FS at $k_F$
 and the ``shadow'' FS  at $k = (\pi,\pi) - k_F$.
 This behavior is at least qualitatively consistent with
 recent ARPES experiments aimed to verify the existence of the outer side of a pocket.~\cite{Zhou} We also emphasize that, as long as SDW order is weak, the incoherent pseudogap peak is the dominant feature in the ARPES spectrum.  This last observation is particularly relevant to LBCO near $1/8$ doping, for which there are indications of SDW ordering. This SDW order is in any case smaller than the order at half-filling, and it is very likely that the dominant ARPES intensity is remains the preudogap despite the potential apearance of SDW order. This would be consistent with the observations in Refs. \onlinecite{he,ton}.

 The redistribution of the fermionic spectral weight in turn affects the pairing problem: in  the presence of SDW precursors,
$T_c$ decreases with decreasing doping~\cite{Sachdev1} because now the pairing is due to the exchange of propagating magnons rather than overdamped spin fluctuations, and the electron-magnon vertex gets smaller as SDW precursors get stronger.~\cite{rep,oleg,schr}
A closely related effect which also reduces $T_c$ is the removal of the low-energy spectral weight due to pseudogap opening~\cite{Kyung:2003}

The precursors to SDW  also give rise to dome-like behavior of the onset temperature for the pairing in electron-doped cuprates (left-hand side of Fig.  \ref{phase-space}) but  there the effect is weaker simply because pairing correlations are weaker.\cite{Armitage09,krotkov}

Recently we  became aware of a
 study of thermal SDW fluctuations  by M. Khodas and A.M. Tsvelik (Ref. \cite{khodas}). Their and our results are similar (e.g., both give linear in $T$
 width of the quasiparticle peaks at low $T$), but not identical as we studied
 isotropic quasi-2D syatems with a true long-range SDW order below a certain
 small $T$ ($\beta \leq 1$) and exponential behavior of the correlation length
 at larger $T$, while Khodas and Tsvelik considered a 2D system with an easy-plane anisotropy and put special emphasis to the fact that spin correlations decay by a power-law at small $T$.

We acknowledge helpful discussions with J-C. Campuzano,
L. Glazman, B. Keimer, E.G. Moon, M. Metlitski, M. Norman, M. Rice, M. V. Sadovskii, S. Sachdev,
 J. Schmalian, O. Sushkov, A. Tsvelik, Z. Tesanovic and A.-M. S. Tremblay.
We are particularly thankful to M.V. Sadovskii and A.-M.S. Tremblay for careful reading of the manuscript and useful comments.
 The work was supported by  NSF-DMR-0906953 (A. V. Ch., T. A. S.) and
NSF  DMR-0847224 (T. A. S.).

\end{document}